\documentclass[sigconf,screen]{acmart}
\AtBeginDocument{%
  }

\setcopyright{acmlicensed}
\copyrightyear{2026}
\acmYear{2026}
\acmDOI{XXXXXXX.XXXXXXX}
\acmConference[CHI2026 Workshop on Tools for Thought]{Tools For Thought}{April 16, 2026}{Barcelona, Spain}

\usepackage[inkscapelatex=false]{svg}
\usepackage{booktabs}
\usepackage{tabularx}
\usepackage{algpseudocode}
\usepackage{algorithm}

\begin{document}

\title{Supporting Reflection and Forward-Looking Reasoning With Data-Driven Questions}

\author{Simon W.S. Fischer}
\email{simon.fischer@donders.ru.nl}
\orcid{0000-0003-2992-6563}
\affiliation{%
  \institution{Donders Institute for Brain, Cognition, and Behaviour, Radboud University}
  \department{Dpt. of Human-Centred Intelligent Systems}
  \city{Nijmegen}
  \country{The Netherlands}
}

\author{Hanna Schraffenberger}
\email{hanna.schraffenberger@ru.nl}
\orcid{0000-0003-1847-2754}
\affiliation{%
  \institution{Interdisciplinary Hub for Digitalization and Society (iHub) and Institute for Computing and Information Sciences (iCIS), Radboud University}
  \city{Nijmegen}
  \country{The Netherlands}
}

\author{Serge Thill}
\email{serge.thill@donders.ru.nl}
\orcid{0000-0003-1177-4119}
\affiliation{%
  \institution{Donders Institute for Brain, Cognition, and Behaviour, Radboud University}
  \department{Dpt. of Human-Centred Intelligent Systems}
  \city{Nijmegen}
  \country{The Netherlands}
}

\author{Pim Haselager}
\email{pim.haselager@donders.ru.nl}
\orcid{0000-0002-4077-9560}
\affiliation{%
  \institution{Donders Institute for Brain, Cognition, and Behaviour, Radboud University}
  \department{Dpt. of Human-Centred Intelligent Systems}
  \city{Nijmegen}
  \country{The Netherlands}
}

\renewcommand{\shortauthors}{Fischer et al.}

\begin{abstract}
  Many generative AI systems as well as decision-support systems (DSSs) provide operators with predictions or recommendations. 
  Various studies show, however, that people can mistakenly adopt the erroneous results presented by those systems. Hence, it is crucial to promote critical thinking and reflection during interaction. One approach we are focusing on involves encouraging reflection during machine-assisted decision-making by presenting decision-makers with data-driven questions. In this short paper, we provide a brief overview of our work in that regard, namely: 1) the development of a question taxonomy, 2) the development of a prototype in the medical domain and the feedback received from clinicians, 3) a method for generating questions using a large language model, and 4) a proposed scale for measuring cognitive engagement in human-AI decision-making. In doing so, we contribute to the discussion about the design, development, and evaluation of tools for thought, i.e., AI systems that provoke critical thinking and enable novel ways of sense-making.
\end{abstract}



\received{March 26, 2026}

\maketitle

\section{Introduction}

Generative AI as well as machine learning models are increasingly being used in decision-making by providing data-driven insights.
Various studies show, however, that people tend to rely too much on the outputs or recommendations of these systems. A study by Anthropic, for example, identifies ``disempowerment potential'' of large language models where people tend to adopt the beliefs embedded in the language model \cite{Sharma2026}. Similarly, it has been shown that language models can influence moral judgement \cite{Krugel2023} and opinions \cite{Jakesch2023}. In machine-assisted decision-making with so-called decision-support systems (DSSs), it has been found that even expert decision-makers, such as clinicians, are prone to accepting wrong recommendations from these systems \cite{Bansal2021, Dratsch2023, Jacobs2021a}. This phenomenon is known as overreliance \cite{Passi2024}. 

Particularly in high-risk sectors, such as healthcare, it is crucial that decision-makers, e.g., clinicians, are supported in such a way that they know when to accept a machine recommendation and when not to. Accordingly, legislation such as the European AI Act mandates \textit{human oversight}, which requires systems to be designed and developed in such a way that persons are enabled  ``to remain aware of the possible tendency of automatically relying or over-relying on the output produced by a high-risk AI system (automation bias)'' (Article 14, 4b).
It is therefore important, and in some cases even necessary, to promote cognitive engagement and continuous critical thinking of the decision-maker or person interacting with the system.

So-called \textit{tools for thought} (TfT) aim to promote critical thinking and reflection. Reflection helps to examine information and scrutinise assumptions, and thus has the potential to improve reasoning and judgement \cite{Khoshgoftar2023, Walger2016, Prakash2019}. One way to promote reflection is through questions. 
Currently, the use of question in the context of human-AI decision-making is relatively underexplored. Related work has been done by \citet{Reicherts2022} who implemented a chatbot to probe people's thinking with context-dependent questions. Another study found that phrasing causal explanations as questions can help people to better assess the validity of statements \cite{Danry2023}. Moreover, in the field of education, chatbots are proposed that utilise the Socratic questioning technique, a common method in education \cite{Naeem2025,Liu2024b,Favero2024}. 
Despite these approaches, many AI systems for decision-making focus on providing answers or advice in the form of predictions, recommendations and explanations.

Explanations help to understand how the system arrived at an outcome. In some cases, however, explanations can increase overreliance, for example if the AI recommendation is incorrect \cite{Bansal2021,Wang2021}. Furthermore, some operators may not be concerned with looking \textit{inside} the ``black-box'' and trying to understand how the system works \cite{Zednik2019}, but would rather like support in the broader decision-making context \cite{Liao2020}. Explanations represent a backward-reasoning from the end result back to the input data \cite{Zhang2024a}. Instead, decision-makers should be promoted in a forward-looking manner \cite{Zhang2024a,Fischer2025}. This includes helping the decision-maker to develop their own line of reasoning, evaluate different options \cite{Miller2023}, and consider the consequences of the decision.

Questions can promote forward-looking reasoning, as they do not provide an answer and allow the person to think for themselves and form their own judgement, i.e., decision self-efficacy. Furthermore, questions can help to make implicit assumptions more explicit, thereby increasing awareness of the reasons behind a particular decision. The ability to recognise and articulate reasons makes it possible, on the one hand, to justify actions retrospectively and thus to be accountable (backward-looking), and more importantly, to direct actions in such a way that a desired state is achieved, thereby taking forward-looking responsibility \cite{Fischer2025}.

Against this background, we therefore propose using questions to support the decision-making process. In the following we will briefly outline the various steps we have taken in this regard, aiming to contribute to the discussion on the design, development, and evaluation of tools for thought:
\begin{itemize}
    \item A taxonomy for identifying relevant elements in human-AI decision-making, to which questions can relate (section~\ref{sect:taxonomy}).
    \item A prototype in the medical field (section~\ref{sect:prototype}) with preliminary feedback from clinicians (section~\ref{sect:study}).
    \item A method for generating data-driven questions using a large language model (section~\ref{sect:llm}).
    \item A new proposed self-report scale for measuring cognitive engagement in human-AI decision-making (section~\ref{sect:scale}).
\end{itemize}

The question taxonomy and the scale for measuring cognitive engagement can be found in the references provided \cite{Fischer2025, Fischer2025b}. The prototype including clinicians' experiences and the method for generating questions with LLMs are currently work in progress \cite{Fischer2026,Fischer2026a}. 

\section{A Taxonomy for Identifying and Formulating Relevant Questions}
\label{sect:taxonomy}

In order to identify and formulate relevant questions that stimulate critical reflection, we have created a taxonomy of elements in human-AI decision-making to which questions can relate to \cite{Fischer2025}. To this end, we adapted a taxonomy of Socratic questions \cite{Paul2019} and mapped it to the context of human-AI decision-making using a question bank for explainable AI \cite{Liao2020}. We also used Bloom's taxonomy to argue that questions should address higher-order cognitive processes. 

We identify 10 dimensions or elements that questions can relate to, such as model behaviour and decision boundaries (e.g., \textit{Would you suggest the same treatment if the patient were 5 years older?}), assumptions or cognitive biases of the decision-maker (e.g., \textit{How does the machine recommendation compare to your assumptions?}), or relevance of data points (e.g., \textit{Is factor \textit{x} relevant to focus on?}). These questions are aimed to help the decision-maker to reconsider the information relating to the patient's case, understand how the DSS works, and reflect on the decision at hand.

While creating the taxonomy, we focused on decision-makers who operate a DSS. Hence, the taxonomy aims to help formulate questions that promote the cognitive engagement among decision-makers. Nevertheless, other stakeholders like developers could also benefit from the taxonomy, by questioning assumptions or the appropriateness of a data set. Furthermore, the taxonomy could be used as a means of promoting AI literacy, for example by helping people to remain more critical of AI-generated information.

\section{Questions in Clinical Decision-Making}
\label{sect:eval}

\subsection{Prototype Design}
\label{sect:prototype}

\begin{figure*}
    \centering
    \includegraphics[width=.65\linewidth]{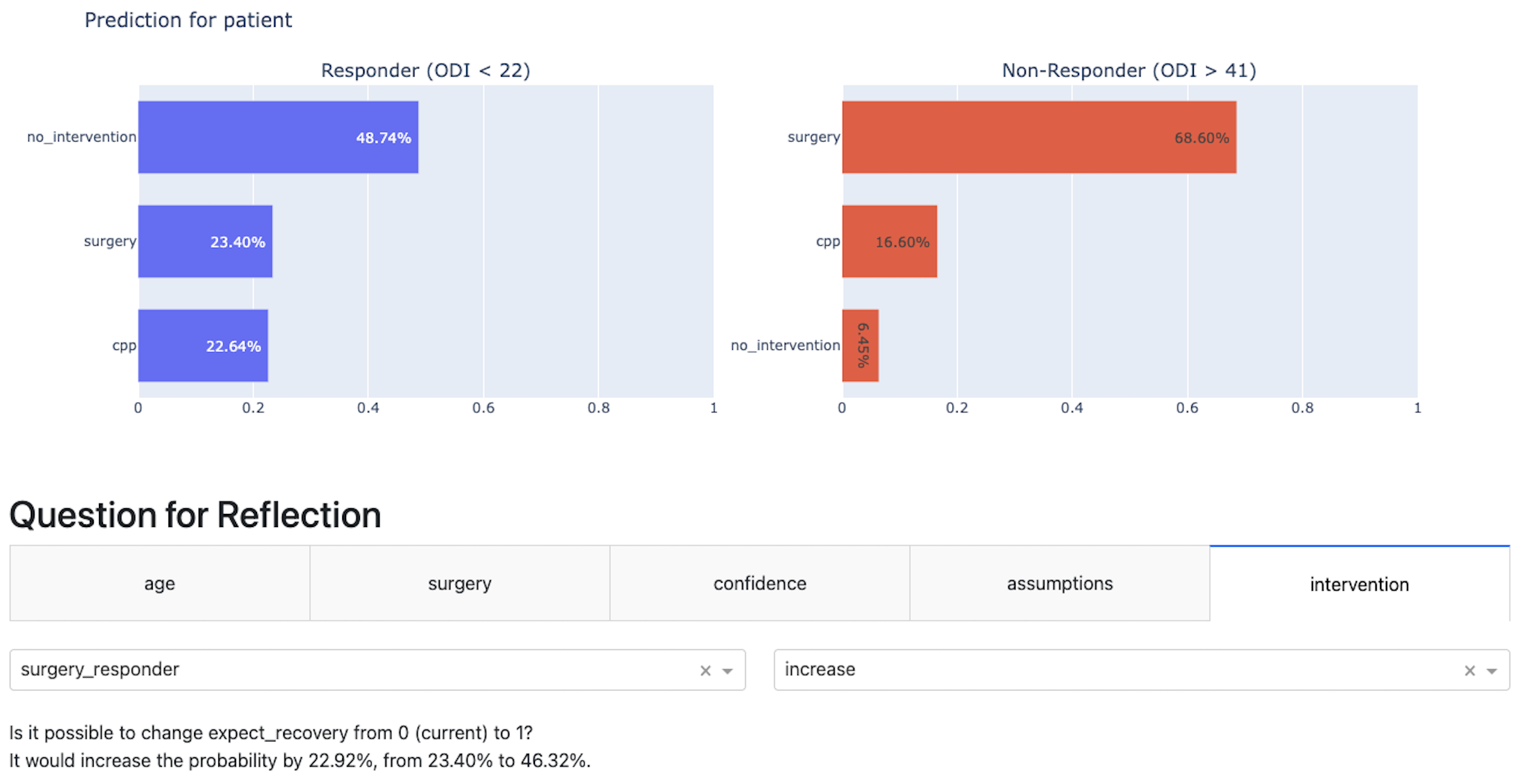}
    \caption{The interface of our prototype: The bar charts show predictions of the effectiveness of three possible treatment options, divided into responder (success) and non-responder (failure) categories. At the bottom, reflective questions are displayed in plain text. The current tab shows the question about the possibility to change an input feature to make the effectiveness of a treatment option more likely (minimum change in input with the maximum effect in outcome). This counterfactual thinking can provide insights into the workings of the DSS. In addition, it could provide an opportunity to first consider other intervention options, such as therapy, and only then consider other treatments, such as surgery.}
    \label{fig:prototype}
    \Description{}
\end{figure*}

We implemented a web-based prototype for a medical decision-task, namely the treatment of chronic low back pain, to assess the feasibility of a system that generates questions and to evaluate its perceived usefulness in the decision-making process \cite{Fischer2026}. For this, we replicated a decision-support system that is used in clinical practice for several years to help clinicians make decisions about treatment options. Importantly, we added the functionality of outputting five different questions based on patient information, the DSS behaviour, and its predictions. Accordingly, the interface shows a question alongside the treatment prognoses (Fig.~\ref{fig:prototype}).

To identify possible questions our system could output, we drew on the question taxonomy described above. We inspected the model behaviour and its decision boundaries through feature perturbation in order to extract information that can be questioned. One question, for example, addresses a hypothetical scenario, similar to counterfactual explanations, and asks about the feasibility of changing certain factors. For this, alternative predictions are calculated by considering all possible feature combinations. The operator then selects, as shown in Fig.~\ref{fig:prototype}, the desired treatment option (first drop-down) and whether the predicted effectiveness for that treatment option should be increased or decreased (second drop-down). Then, the alternative prediction is selected which, compared to the current prediction, shows the smallest change in input features, i.e., patient information, and the greatest effect on the outcome, i.e., treatment prediction. In addition, we also implemented a general question that is not informed by data. The five questions are listed in Table~\ref{tab:questions}.

\begin{table}[]
\caption{The five questions that our prototype outputs alongside the treatment recommendation. We used our question taxonomy to identify relevant questions.}
\label{tab:questions}
\begin{tabularx}{\columnwidth}{lX}
\toprule
\textbf{Taxonomy ID} &
\textbf{Question}  \\ 
\midrule
Q10 &
Is the patient’s age of 47 years relevant to consider in this case? The patient of 47 years is approaching the red-flag threshold which is 50 years. Age correlates with the effect of surgery. \\
Q1 &
When was the specified surgery performed and at which location of the spine? Previous surgeries reduce the effect of surgery by 15\% - 25\%. \\
Q6 &
How confident are you about your decision? The confidence of the prediction for the most effective treatment (surgery 59.92\%) is at 42.58\%. \\
Q6 & 
Does the prediction change your initial judgement? If so, why? \\
Q9 & 
Is it possible to change the patient's \textit{expected recovery} from `no' (current) to `yes'? It would increase the expected effectiveness of surgery by 22.92\%, from 23.40\% to 46.32\%. \\
\bottomrule
\end{tabularx}
\end{table}

\subsection{Preliminary Findings: Perceptions of Clinicians}
\label{sect:study}

We presented our prototype to clinical spine experts (n=6) and asked them about their perceptions in semi-structured, in-situ interviews \cite{Fischer2026}. We recorded and transcribed the interviews and analysed the data using thematic analysis to gain initial insights into our two research questions:

\begin{itemize}
    \item \textbf{RQ1}. How do clinicians perceive questions during decision-making?
    \item \textbf{RQ2}. What makes questions effective from a clinician’s perspective?
\end{itemize}

The participants reported a high familiarity with our prototype, since the functionality of our prototype and the visual representation of the treatment predictions are based on the DSS they use in clinical practice. Regarding \textit{RQ1}, we identified five themes, namely 1) questions can provide insights into how the DSS works, 2) questions can function as
reminders to check information, 3) questions can prompt discussions with patients, 4) questions can help to consider alternatives to the most probable option, and 5) questions are perceived differently depending on the person.

The general sentiment in the interviews was that the questions presented in our prototype would not be particularly helpful to clinicians in reflecting on their decision-making. This is because, as the participants noted, certain factors (i.e., features) referred to in the questions were either part of their standard reasoning (e.g., previous surgeries), or were considered less crucial (e.g., patient's age). Nevertheless, the clinicians also reported that questions could serve as reminders to check or consider certain information, particularly when one is less focused, such as at the end of the day, or that questions could help novice clinicians better understand how the DSS works, thereby preventing them from simply relying on the DSS prediction. For example, although the DSS takes into account the patient's previous surgeries, the length of time since the operation is significant. A previous surgery that has a great impact on the DSS prediction may be considered negligible if it was performed 20 years ago. 

Although we have found that participants perceive the usefulness of a particular question differently, we identified five characteristics that make questions more effective (\textit{RQ2}), namely: 1) Questions must fit the context, 2) questions must address factors that are not part of standard reasoning, 3) questions must be answerable and actionable, 4) the timing of questions matters, and 5) questions should not be too demanding.

\section{Generating Questions with a Language Model}
\label{sect:llm}

\begin{figure}
    \centering
    \includegraphics[width=1\linewidth]{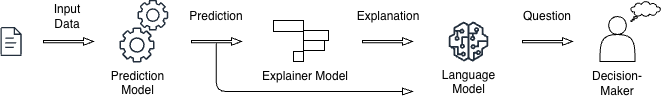}
    \caption{A flowchart illustrating our question generation system. Based on input data, such as patient information, a decision-support system computes a prediction. An explanation is then generated for this prediction in the form of feature contribution (LIME). Finally, the explanation and the prediction are passed to a language model to formulate a question. The generated question helps the decision-maker reflect on the prediction and decision at hand.}
    \label{fig:pipeline}
    \Description{}
\end{figure}

The initial questions in our current prototype are predefined. To make the generation of questions more flexible and generalisable, we propose an approach that utilises a large language model \cite{Fischer2026a}. More specifically, we take the DSS prediction, generate an explanation in the form of feature contribution through LIME, and prompt a local language model with this information (Fig.~\ref{fig:pipeline}). Our prompts are designed to generate questions that align with our question taxonomy. 

As such, the feature contribution, i.e., LIME explanation, allows to generate questions that relate, for example, to the most important feature and its relevance (taxonomy ID Q2), or to the features with a negative contribution, asking why the option should be considered despite the factors that speak against it (taxonomy ID Q4). 
An example prompt could be: \textit{``The most effective treatment is likely to be [$prediction$]. The features for this prediction are [$positive\_features$], while the features against this prediction are [$negative\_features$]. Formulate a question that stimulates the decision-maker to reflect on the prediction, asking why they would go for this decision, despite the reasons against it. Keep the question concise.''}. A resulting example question is: \textit{``Why would you prioritize conservative care despite concerns about recovery expectation, prior surgery, and neuromuscular conditions?''}. 

With this approach, which is related to natural language explanations in the field of human-centred explainable AI, it is possible to translate technical explanations in the form feature contribution, which might be difficult to understand for decision-makers, into a question that promotes critical thinking.

We noticed, however, that a small amount of the LLM-generated questions contained variables, e.g., hemoglobin levels, that do not appear in our dataset of patient cases and are thus not listed in the explanation, i.e., the feature
contributions. This should remind us that LLMs are statistical models that generate text based on patterns derived from the data they were trained on. To minimise the risk of generating irrelevant or incorrect questions, which in the worst case could be misleading, it would be possible to implement a RAG (retrieval-augmented generation) approach in order
to add contextual or domain-specific knowledge.

\section{A Scale for Measuring Cognitive Engagement}
\label{sect:scale}

Current evaluation methods of decision-support systems or other interventions to mitigate overreliance focus on the decision accuracy, i.e., whether the correct decision was made. We argue, however, that evaluation of tools for thought must take into account the decision-making process. As such, it should be considered whether the TfT helped to consider alternatives, or to understand the available options.  
We therefore propose possible items for a new self-report scale designed to measure cognitive engagement in human-AI decision-making, and thus the effectiveness of tools for thought such as our question-generation system \cite{Fischer2025b}. Accordingly, example items of a scale could include, \textit{``The system (TfT) helped me to be aware of my preferences/assumptions''}, or \textit{``The system (TfT) helped me to compare and contrast different options''}. 

We have derived and adapted potential items from existing scales used in education to assess how students use technology to perform cognitive learning activities, as well as to measure associated factors to cognitive engagement, such as decision self-efficacy. We have not used this scale in our exploratory evaluation of our prototype (section~\ref{sect:eval}), as we developed the potential scale in parallel with the first feedback sessions with the clinicians. Future work is necessary to develop a validated scale. 

In addition to assessing the impact of generative AI on critical thinking and reflection, such a scale could serve as a guide for the design and development of future tools for thought aimed at promoting cognitive engagement.

\section{Concluding Remarks}
\label{sect:future-work}

Some open questions remain for future work. Namely, how to formulate relevant questions in more open-ended human-AI interactions. Generating relevant questions might be more difficult once there is no decision-support system with explainable AI, i.e., underlying logic/structure how input data connects to an outcome.

Moreover, as our exploratory study suggests, questions can be perceived differently by different people. Therefore it will be important for various tools for thought, such as our question-generation system, to adapt to individual decision-makers. People differ in their decision-making styles, their need for cognition, and their risk attitudes -- all factors that influence decision-making. Moreover, external factors like time constraints also play a role. Future work would need to investigate how these aspects can be taken into account in order to generate relevant questions. Our proposed approach of using a language model could serve as a first step towards adapting and personalising questions. 

Since reflection has been shown to improve decision-making, it seems promising to utilise generative AI in this regard to promote cognitive engagement, forward reasoning, and decision self-efficacy.

\begin{acks}
This research is funded by the Donders Centre for Cognition. We thank two anonymous reviewers for their constructive feedback.
\end{acks}

\bibliographystyle{ACM-Reference-Format}
\bibliography{main}

@article{Dratsch2023,
  title = {Automation {{Bias}} in {{Mammography}}: {{The Impact}} of {{Artificial Intelligence}}                     {{BI-RADS Suggestions}} on {{Reader Performance}}},
  shorttitle = {Automation {{Bias}} in {{Mammography}}},
  author = {Dratsch, Thomas and Chen, Xue and Rezazade Mehrizi, Mohammad and Kloeckner, Roman and {M{\"a}hringer-Kunz}, Aline and P{\"u}sken, Michael and Bae{\ss}ler, Bettina and Sauer, Stephanie and Maintz, David and {Pinto dos Santos}, Daniel},
  year = 2023,
  month = may,
  journal = {Radiology},
  volume = {307},
  number = {4},
  pages = {e222176},
  issn = {0033-8419, 1527-1315},
  doi = {10.1148/radiol.222176},
  urldate = {2024-09-11},
  langid = {english},
}

@techreport{Passi2024,
  title = {Appropriate Reliance on Generative {{AI}}: {{Research}} Synthesis},
  author = {Passi, Samir and Dhanorkar, Shipi and Vorvoreanu, Mihaela},
  year = 2024,
  month = mar,
  number = {MSR-TR-2024-7},
  institution = {Microsoft},
}

@inproceedings{Danry2023,
  title = {Don't {{Just Tell Me}}, {{Ask Me}}: {{AI Systems}} That {{Intelligently Frame Explanations}} as {{Questions Improve Human Logical Discernment Accuracy}} over {{Causal AI}} Explanations},
  shorttitle = {Don't {{Just Tell Me}}, {{Ask Me}}},
  booktitle = {Proceedings of the 2023 {{CHI Conference}} on {{Human Factors}} in {{Computing Systems}}},
  author = {Danry, Valdemar and Pataranutaporn, Pat and Mao, Yaoli and Maes, Pattie},
  year = 2023,
  month = apr,
  series = {{{CHI}} '23},
  pages = {1--13},
  publisher = {ACM},
  address = {Hamburg Germany},
  doi = {10.1145/3544548.3580672},
  urldate = {2023-04-25},
  isbn = {978-1-4503-9421-5},
  langid = {english},
}

@inproceedings{Wang2021,
  title = {Are {{Explanations Helpful}}? {{A Comparative Study}} of the {{Effects}} of {{Explanations}} in {{AI-Assisted Decision-Making}}},
  shorttitle = {Are {{Explanations Helpful}}?},
  booktitle = {26th {{International Conference}} on {{Intelligent User Interfaces}}},
  author = {Wang, Xinru and Yin, Ming},
  year = 2021,
  month = apr,
  pages = {318--328},
  publisher = {ACM},
  address = {College Station TX USA},
  doi = {10.1145/3397481.3450650},
  urldate = {2024-09-16},
  isbn = {978-1-4503-8017-1},
  langid = {english},
}

@inproceedings{Miller2023,
  title = {Explainable {{AI}} Is {{Dead}}, {{Long Live Explainable AI}}!: {{Hypothesis-driven Decision Support}} Using {{Evaluative AI}}},
  shorttitle = {Explainable {{AI}} Is {{Dead}}, {{Long Live Explainable AI}}!},
  booktitle = {2023 {{ACM Conference}} on {{Fairness}}, {{Accountability}}, and {{Transparency}}},
  author = {Miller, Tim},
  year = 2023,
  month = jun,
  series = {{{FAccT}} '23},
  pages = {333--342},
  publisher = {ACM},
  address = {Chicago IL USA},
  doi = {10.1145/3593013.3594001},
  urldate = {2024-07-18},
  isbn = {979-8-4007-0192-4},
  langid = {english},
}

@article{Jacobs2021a,
  title = {How Machine-Learning Recommendations Influence Clinician Treatment Selections: The Example of Antidepressant Selection},
  shorttitle = {How Machine-Learning Recommendations Influence Clinician Treatment Selections},
  author = {Jacobs, Maia and Pradier, Melanie F. and McCoy, Thomas H. and Perlis, Roy H. and {Doshi-Velez}, Finale and Gajos, Krzysztof Z.},
  year = 2021,
  month = feb,
  journal = {Translational Psychiatry},
  volume = {11},
  number = {1},
  pages = {108},
  issn = {2158-3188},
  doi = {10.1038/s41398-021-01224-x},
  urldate = {2024-08-09},
  langid = {english},
}

@article{Prakash2019,
  title = {Interventions to Improve Diagnostic Decision Making: {{A}} Systematic Review and Meta-Analysis on Reflective Strategies},
  shorttitle = {Interventions to Improve Diagnostic Decision Making},
  author = {Prakash, Shivesh and Sladek, Ruth M. and Schuwirth, Lambert},
  year = 2019,
  journal = {Medical Teacher},
  volume = {41},
  number = {5},
  pages = {517--524},
  issn = {0142-159X, 1466-187X},
  doi = {10.1080/0142159X.2018.1497786},
  urldate = {2025-01-04},
  langid = {english},
}

@article{Walger2016,
  title = {{{HR}} Managers' Decision-Making Processes: A ``Reflective Practice'' Analysis},
  shorttitle = {{{HR}} Managers' Decision-Making Processes},
  author = {Walger, Carolina and Roglio, Karina De Dea and Abib, Gustavo},
  year = 2016,
  month = jun,
  journal = {Management Research Review},
  volume = {39},
  number = {6},
  pages = {655--671},
  issn = {2040-8269},
  doi = {10.1108/MRR-11-2014-0250},
  urldate = {2025-01-06},
  copyright = {https://www.emerald.com/insight/site-policies},
  langid = {english},
}

@article{Khoshgoftar2023,
  title = {Medical Students' Reflective Capacity and Its Role in Their Critical Thinking Disposition},
  author = {Khoshgoftar, Zohreh and {Barkhordari-Sharifabad}, Maasoumeh},
  year = 2023,
  month = mar,
  journal = {BMC Medical Education},
  volume = {23},
  number = {1},
  pages = {198},
  issn = {1472-6920},
  doi = {10.1186/s12909-023-04163-x},
  urldate = {2025-01-06},
  langid = {english},
}

@inproceedings{Bansal2021,
  title = {Does the {{Whole Exceed}} Its {{Parts}}? {{The Effect}} of {{AI Explanations}} on {{Complementary Team Performance}}},
  shorttitle = {Does the {{Whole Exceed}} Its {{Parts}}?},
  booktitle = {Proceedings of the 2021 {{CHI Conference}} on {{Human Factors}} in {{Computing Systems}}},
  author = {Bansal, Gagan and Wu, Tongshuang and Zhou, Joyce and Fok, Raymond and Nushi, Besmira and Kamar, Ece and Ribeiro, Marco Tulio and Weld, Daniel},
  year = 2021,
  month = may,
  pages = {1--16},
  publisher = {ACM},
  address = {Yokohama Japan},
  doi = {10.1145/3411764.3445717},
  urldate = {2023-11-29},
  isbn = {978-1-4503-8096-6},
  langid = {english},
}

@inproceedings{Jakesch2023,
  title = {Co-{{Writing}} with {{Opinionated Language Models Affects Users}}' {{Views}}},
  booktitle = {Proceedings of the 2023 {{CHI Conference}} on {{Human Factors}} in {{Computing Systems}}},
  author = {Jakesch, Maurice and Bhat, Advait and Buschek, Daniel and Zalmanson, Lior and Naaman, Mor},
  year = 2023,
  month = apr,
  pages = {1--15},
  publisher = {ACM},
  address = {Hamburg Germany},
  doi = {10.1145/3544548.3581196},
  urldate = {2023-10-13},
  isbn = {978-1-4503-9421-5},
  langid = {english},
}

@misc{Sharma2026,
  title = {Who's in {{Charge}}? {{Disempowerment Patterns}} in {{Real-World LLM Usage}}},
  shorttitle = {Who's in {{Charge}}?},
  author = {Sharma, Mrinank and McCain, Miles and Douglas, Raymond and Duvenaud, David},
  year = 2026,
  month = jan,
  number = {arXiv:2601.19062},
  eprint = {2601.19062},
  primaryclass = {cs},
  publisher = {arXiv},
  doi = {10.48550/arXiv.2601.19062},
  urldate = {2026-02-09},
  archiveprefix = {arXiv},
}

@inproceedings{Reicherts2022,
    address = {Glasgow United Kingdom},
    author = {Reicherts, Leon and Park, Gun Woo and Rogers, Yvonne},
    booktitle = {Proceedings of the 4th {{Conference}} on {{Conversational User Interfaces}}},
    doi = {10.1145/3543829.3543832},
    isbn = {978-1-4503-9739-1},
    langid = {english},
    month = {July},
    pages = {1--10},
    publisher = {ACM},
    shorttitle = {Extending {{Chatbots}} to {{Probe Users}}},
    title = {Extending {{Chatbots}} to {{Probe Users}}: {{Enhancing Complex Decision-Making Through Probing Conversations}}},
    urldate = {2025-05-26},
    year = {2022}
}

@article{Naeem2025,
    author = {Naeem, Hadeel},
    doi = {10.1017/epi.2025.10089},
    issn = {1742-3600, 1750-0117},
    journal = {Episteme},
    langid = {english},
    month = {November},
    pages = {1--18},
    title = {Teaching {{Skills}} and {{Intellectual Virtues}} with {{Generative AI}}},
    urldate = {2025-11-13},
    year = {2025}
}

@inproceedings{Liu2024b,
    author = {Liu, Jiayu and Huang, Zhenya and Xiao, Tong and Sha, Jing and Wu, Jinze and Liu, Qi and Wang, Shijin and Chen, Enhong},
    booktitle = {The {{Thirty-eighth Annual Conference}} on {{Neural Information Processing Systems}}},
    langid = {english},
    pages = {85693--85721},
    publisher = {Curran Associates, Inc.},
    title = {{{SocraticLM}}: {{Exploring Socratic Personalized Teaching}} with {{Large Language Models}}},
    volume = {37},
    year = {2024}
}

@misc{Favero2024,
    archiveprefix = {arXiv},
    author = {Favero, Lucile and {P{\'e}rez-Ortiz}, Juan Antonio and K{\"a}ser, Tanja and Oliver, Nuria},
    eprint = {2409.05511},
    langid = {english},
    month = {September},
    number = {arXiv:2409.05511},
    primaryclass = {cs},
    publisher = {arXiv},
    title = {Enhancing {{Critical Thinking}} in {{Education}} by Means of a {{Socratic Chatbot}}},
    urldate = {2024-11-19},
    year = {2024}
}

@article{Zhang2024a,
    author = {Zhang, Zelun Tony and Feger, Sebastian S. and Dullenkopf, Lucas and Liao, Rulu and S{\"u}sslin, Lou and Liu, Yuanting and Butz, Andreas},
    doi = {10.1145/3687024},
    issn = {2573-0142},
    journal = {Proceedings of the ACM on Human-Computer Interaction},
    langid = {english},
    month = {November},
    number = {CSCW2},
    pages = {1--32},
    shorttitle = {Beyond {{Recommendations}}},
    title = {Beyond {{Recommendations}}: {{From Backward}} to {{Forward AI Support}} of {{Pilots}}' {{Decision-Making Process}}},
    urldate = {2025-04-01},
    volume = {8},
    year = {2024}
}

@article{Fischer2025,
    author = {Fischer, Simon W.S. and Schraffenberger, Hanna and Thill, Serge and Haselager, Pim},
    doi = {10.1609/aies.v8i1.36602},
    issn = {3065-8365},
    journal = {Proceedings of the AAAI/ACM Conference on AI, Ethics, and Society},
    month = {October},
    number = {1},
    pages = {940--954},
    title = {A {{Taxonomy}} of {{Questions}} for {{Critical Reflection}} in {{Machine-Assisted Decision-Making}}},
    urldate = {2025-10-30},
    volume = {8},
    year = {2025}
}

@inproceedings{Fischer2025b,
  title = {How to {{Measure Cognitive Engagement}} in {{Machine-Assisted Decision-Making}}?},
  booktitle = {Proceedings of the {{Workshops}} at the {{Fourth International Conference}} on {{Hybrid Human-Artificial Intelligence}} Co-Located with the {{Fourth International Conference}} on {{Hybrid Human-Artificial Intelligence}} ({{HHAI}} 2025)},
  author = {Fischer, Simon W.S. and Schraffenberger, Hanna},
  year = 2025,
  volume = {4074},
  pages = {98--109},
  publisher = {CEUR Workshop Proceedings},
  address = {Pisa, Italy},
  langid = {english},
}

@misc{Fischer2026,
  title = {Questions {{Can Provide Insights}} and {{Function}} as {{Reminders}}: {{Clinicians}}' {{Perceptions}} of {{A Decision-Support System That Presents Data-Driven Questions}}},
  author = {Fischer, Simon W.S. and Schraffenberger, Hanna and {van Hooff}, Miranda and Thill, Serge and Haselager, Pim},
  year = 2026,
  note = {Manuscript submitted for publication.},
  archiveprefix = {},
  langid = {english},
}

@misc{Fischer2026a,
  title = {From {{Explanations}} to {{Questions}}: {{Adding Friction}} and {{Promoting Cognitive Engagement}} in {{Human-AI Decision-Making}}},
  author = {Fischer, Simon W.S. and Holmberg, Linus and Thill, Serge and Schraffenberger, Hanna},
  year = 2026,
  note = {Manuscript submitted for publication.},
  archiveprefix = {},
  langid = {english},
}

@article{Krugel2023,
  title = {{{ChatGPT}}'s Inconsistent Moral Advice Influences Users' Judgment},
  author = {Kr{\"u}gel, Sebastian and Ostermaier, Andreas and Uhl, Matthias},
  year = 2023,
  month = apr,
  journal = {Scientific Reports},
  volume = {13},
  number = {1},
  pages = {4569},
  issn = {2045-2322},
  doi = {10.1038/s41598-023-31341-0},
  urldate = {2024-10-06},
  langid = {english},
}

@book{Paul2019,
    address = {Blue Ridge Summit},
    author = {Paul, Richard and Elder, Linda},
    isbn = {978-0-944583-31-9 978-1-5381-3381-1},
    langid = {english},
    publisher = {Rowman \& Littlefield Publishers},
    series = {Thinker's {{Guide Library}}},
    title = {The {{Thinker}}'s {{Guide}} to {{Socratic Questioning}}},
    year = {2019}
}

@inproceedings{Liao2020,
    address = {Honolulu HI USA},
    author = {Liao, Q. Vera and Gruen, Daniel and Miller, Sarah},
    booktitle = {Proceedings of the 2020 {{CHI Conference}} on {{Human Factors}} in {{Computing Systems}}},
    doi = {10.1145/3313831.3376590},
    isbn = {978-1-4503-6708-0},
    langid = {english},
    month = {April},
    pages = {1--15},
    publisher = {ACM},
    series = {{{CHI}} '20},
    shorttitle = {Questioning the {{AI}}},
    title = {Questioning the {{AI}}: {{Informing Design Practices}} for {{Explainable AI User Experiences}}},
    urldate = {2024-07-18},
    year = {2020}
}

@article{Zednik2019,
  title = {Solving the {{Black Box Problem}}: {{A Normative Framework}} for {{Explainable Artificial Intelligence}}},
  shorttitle = {Solving the {{Black Box Problem}}},
  author = {Zednik, Carlos},
  year = 2019,
  month = dec,
  journal = {Philosophy \&amp; Technology},
  volume = {34},
  number = {2},
  pages = {265--288},
  issn = {2210-5433, 2210-5441},
  doi = {10.1007/s13347-019-00382-7},
  urldate = {2020-11-06},
  langid = {english},
}


\end{document}